\documentstyle[12pt,twoside,fleqn,espcrc1,epsfig]{article}

\newcommand{\AmS}{{\protect\the\textfont2
  A\kern-.1667em\lower.5ex\hbox{M}\kern-.125emS}}

\title{
\vspace*{-4cm}
\begin{flushright}
DPNU-99-23\\
KFKI-1999-03/A\\
\vspace*{1.4cm}
\end{flushright}
{\bf Back-to-Back Correlations for Bosons Modified by Medium}\thanks{
This research was partly supported by the
US - Hungarian Joint Fund, the Hungarian OTKA grant
T024094, an OTKA - NWO grant,
the U.S. Department of Energy contracts No. DE-FG02-93ER40764,
DE-FG-02-92-ER40699 and DE-AC02-76CH00016 and the Grant-in-Aid
for Scientific Research No. 10740112 of the Japanese Ministry
of Education, Science and Culture.
}}

\author{M. Asakawa\address{Department of Physics, School of Science,
	Nagoya University, Nagoya 464-8602, Japan},
        T. Cs\"org\H o\address{MTA KFKI RMKI, H - 1525 Budapest 114,
	POB. 49, Hungary},
	and
	M. Gyulassy\address{Dept. of Physics, Columbia University,
        538 W 120th St, New York, NY 10027, U.S.A.}
}

\begin{document}

\maketitle

\pagestyle{empty}

\begin{abstract}
Novel back-to-back correlations are shown to arise
for thermal ensembles of squeezed bosonic states associated 
with medium-modified mass-shifts.
The strength of these correlations could become unexpectedly large
in heavy ion collisions.
\end{abstract}

\section{INTRODUCTION}
In this paper we consider the effect of possible
mass shifts in the dense medium on two boson correlations in general.
Thus far medium modifications of hadron masses have been mainly
considered in terms of effects on such observables as
dilepton yields and spectra.
Hadron mass shifts are caused by interactions
in a dense medium and therefore vanish on the freeze-out surface.
Thus, a naive first expectation is that in medium hadron
modifications may have little or no effect on two boson correlations,
and so the usual Hanbury-Brown Twiss (HBT)
effect \cite{hbt}
has been expected to be only concerned
with the geometry and matter flow gradients on the freeze-out surface.
However, in this paper we show that an interesting
quantum mechanical correlation is induced due to the
fact that medium modified bosons can be represented in terms of
two-mode squeezed states of the asymptotic bosons,
which are observables.

In this paper we assume the validity of relativistic
hydrodynamics up to freeze-out.
The local temperature $T(x)$ and chemical potential $\mu(x)$ are given.
In relativistic heavy ion collisions at CERN SPS, it has been observed
that the one particle spectra \cite{na44}, simultaneously with the
two particle correlation function, can be described by
(local) thermal
distributions fairly precisely~\cite{scl}.
We assume that the sudden approximation
is a valid abstraction in describing
the freeze-out process in relativistic heavy ion collisions
quantum mechanically 
and that there exists an abrupt freeze-out
surface, $\Sigma^\mu(x)$.
        
Let us consider the following model Hamiltonian for a scalar field
$\phi({\bf x})$ in the rest frame of matter,
\begin{equation}
{H} =  H_0 - \frac{1}{2} \int d {\bf x} d {\bf y} \phi({\bf x})
\delta M^2({\bf x}-{\bf y}) \phi({\bf y}), 
\quad
        H_0  = \frac{1}{2} \int d {\bf x} \left(
                \dot{\phi}^2+ |\nabla \phi|^2
                +
                        m_0^2 \phi^2  \right),
\end{equation}
where $H_0$ is the asymptotic Hamiltonian.
The field $\phi({\bf x})$ in $H$ corresponds to 
quasi - particles that propagate with a momentum-dependent
effective mass,
which is related to the vacuum  mass, $m_0$,  via
$m_*^2({|{\bf k}|}) =  m_0^2 - \delta M^2({|{\bf k}|})$.
The mass-shift is assumed to be limited to long wavelength 
collective modes.

The invariant
single-particle and two-particle momentum distributions
are given by
\begin{eqnarray}
N_1({\bf k}_1) & = & \omega_{{\bf k}_1}{d^3N \over d{\bf k}_1} 
        = \omega_{{\bf k}_1} \langle
 a^\dagger_{{\bf k}_1} a^{\phantom\dagger}_{{\bf k}_1}\rangle , \\
N_2({\bf k}_1,{\bf k}_2) & = & 
\omega_{{\bf k}_1} \omega_{{\bf k}_2} 
        \langle a^\dagger_{{\bf k}_1} a^\dagger_{{\bf k}_2} 
        a^{\phantom\dagger}_{{\bf k}_2}
        a^{\phantom\dagger}_{{\bf k}_1}\rangle \nonumber \\
& = &  
\omega_{{\bf k}_1} \omega_{{\bf k}_2}
(\langle a^\dagger_{{\bf k}_1} a^{\phantom\dagger}_{{\bf k}_1}\rangle
\langle  a^\dagger_{{\bf k}_2} a^{\phantom\dagger}_{{\bf k}_2} \rangle + 
\langle a^\dagger_{{\bf k}_1} a^{\phantom\dagger}_{{\bf k}_2}\rangle
\langle  a^\dagger_{{\bf k}_2} a^{\phantom\dagger}_{{\bf k}_1} \rangle + 
\langle a^\dagger_{{\bf k}_1} a^\dagger_{{\bf k}_2}\rangle
\langle  a^{\phantom\dagger}_{{\bf k}_2}
         a^{\phantom\dagger}_{{\bf k}_1} \rangle),
\label{rand}
\end{eqnarray}
where $a_{\bf k}$ is the annihilation operator for the
asymptotic quantum with four-momentum 
$k^{\mu}\, = \, (\omega_{\bf k},{\bf k})$,
$\omega_{\bf k}^2 = {m_0^2 + {\bf k}^2}$  and
the expectation value of an operator $\hat{O}$ is given by
the density matrix $\hat{\rho}$ as 
$\langle \hat{O} \rangle = {\rm Tr} \, \hat{\rho}\, \hat{O}$.

We introduce the 
chaotic and squeezed amplitudes, defined, respectively, as
\begin{equation}
G_c(1,2) = 
\sqrt{\omega_{{\bf k}_1} \omega_{{\bf k}_2} }
  \langle a^{\dagger}_{{\bf k}_1} a^{\phantom\dagger}_{{\bf k}_2}\rangle,
\quad
G_s(1,2) =  
\sqrt{\omega_{{\bf k}_1} \omega_{{\bf k}_2} }
  \langle a^{\phantom\dagger}_{{\bf k}_1}
  a^{\phantom\dagger}_{{\bf k}_2} \rangle .
\label{gs}
\end{equation}
In most situations, the chaotic amplitude, $G_c(1,2) \equiv G(1,2)$
is dominant, and carries the Bose-Einstein correlations,
while the squeezed amplitude, $G_s(1,2)$ vanishes:
\begin{equation}
C_2({\bf k}_1,{\bf k}_2)  =  {N_2({\bf k}_1,{\bf k}_2)
\over N_1({\bf k}_1) N_1({\bf k}_2) } = 
        1 + { | G(1,2) |^2 \over G(1,1) G(2,2) }.
\label{hbt}
\end{equation}
The exact value of the intercept, $C_2({\bf k},{\bf k})=2$,
is a characteristic signature  of a chaotic Bose gas
without dynamical 2-body correlations.

\section{RESULTS FOR A HOMOGENEOUS SYSTEM}
The terms neglected in (\ref{hbt}) involving $G_s(1,2)$
become non-negligible when 
$\delta M^2(|{\bf k}|)\ne 0$. Given such a mass shift,
the dispersion relation is modified to
$\Omega_{\bf k}^2 =\omega^2_{\bf k}-\delta M^2(|{\bf k}|)$,
where $\Omega_{\bf k}$ is the frequency of the in-medium mode
with momentum ${\bf k}$.
The annihilation operator for the in-medium quasi-particle with
momentum ${\bf k}$, $b_{\bf k}$, and that of the asymptotic
field, $a_{\bf k}$, are related by a Bogoliubov
transformation \cite{ac}: 
\begin{equation}
a^{\phantom{\dagger}}_{{\bf k}_1}
        = U^\dagger b^{\phantom{\dagger}}_{{\bf k}_1} U 
        = c^{\phantom{\dagger}}_{{\bf k}_1}
          b^{\phantom{\dagger}}_{{\bf k}_1} + 
                s^{*\phantom{\dagger}}_{-{\bf k}_1} b^\dagger_{-{\bf k}_1}
\equiv C^{\phantom{\dagger}}_1 + S^\dagger_{-1},
\label{asq}
\end{equation}
where
$c_{\bf k}=\cosh[r_{\bf k}]$, 
$s_{\bf k}=\sinh[r_{\bf k}]$, and
$r_{\bf k}=\frac{1}{2}\log(\omega_{\bf k}/\Omega_{\bf k})$.
As is well-known, the Bogoliubov
transformation is equivalent to a squeezing operation, and so
we call $r_{\bf k}$ the mode dependent squeezing parameter.
While it is the $a$-quanta that are observed, it is the $b$-quanta
that are thermalized in medium. Thus, we consider the
thermal average for a globally thermalized gas of 
the $b$-quanta, that is homogeneous in volume $V$:
\begin{equation}
\hat{\rho} = {\displaystyle\phantom{|} 1 \over Z} \exp\left(-\frac{1}{T}
        \frac{V}{(2 \pi)^3} \int d {\bf k}\, \Omega_{\bf  k}
                \, b^\dagger_{\bf k} b^{\phantom\dagger}_{\bf k}\right).
\end{equation}
When this thermal average is applied,
\begin{eqnarray}
N_1({\bf k}) & \!\!\!= \!\!\!& \frac{V}{(2 \pi)^3}\,
\omega_{\bf k}\, n_1({\bf k}),
~~n_1({\bf k})  = 
        | c_{\bf k}^{\phantom{\dagger}}|^2 
        n_{\bf k}^{\phantom{\dagger}} 
        + |  s_{-\bf k}|^2 (n_{-\bf k} + 1),
~~n_{\bf k} = \frac{1}{\exp(\Omega_{\bf k}/T) -1},\nonumber \\
G_c(1,2) & \!\!\!= \!\!\!&
\sqrt{\omega_{{\bf k}_1} \omega_{{\bf k}_2}}
\left[\langle C^\dagger_1C^{\phantom\dagger}_2\rangle + 
\langle
S^{\phantom\dagger}_{-1} S^\dagger_{-2}\rangle\right],
~
G_s(1,2) =
\sqrt{\omega_{{\bf k}_1} \omega_{{\bf k}_2}}
\left[
\langle S^\dagger_{-1} C^{\phantom\dagger}_2
\rangle + \langle C^{\phantom\dagger}_1 S^\dagger_{-2}
\rangle 
\right]\!
. 
\end{eqnarray}
In the homogeneous case, the resulting two particle correlation
function is unity except for the parallel and anti-parallel cases.
The {\em dynamical} correlation due to the two mode squeezing associated
with mass shifts is therefore
{\em back-to-back} as first pointed out in \cite{ac}.
The HBT correlation intercept 
remains 2 for identical momenta. Evaluating 
$C_2({\bf k}, {\bf -k})$
for $T \simeq 140$
MeV, $|{\bf k}| = 0$ -- 500 MeV for $\phi$ mesons,
as a function of $m^*_{\phi}$, one finds 
back-to-back correlations (BBC) as big as 100 -- 1000
for reasonable values of $m^*_{\phi}$.
Note, that these novel BBC-s are not bounded from above,
$1 < 
C_2({\bf k}, {\bf -k}) < \infty $. With increasing 
values of $| {\bf k}|$ these BBC-s increase indefinitely. 
The huge BBC of decaying medium is, however, reduced
if the decay of the medium is not completely sudden.
To describe a more gradual freeze-out, the 
probability distribution $F(t_i)$ of the decay times $t_i$
is introduced. The time evolution of $a_{\bf k}(t)$
is $a_{\bf k}(t) = a_{\bf k}(t_i) \exp[-i \omega_{\bf k} (t - t_i)]$,
which leads to
\begin{equation}
C_2({\bf k}, {\bf -k}) = 1 +  
{\displaystyle\phantom{|} {|c^*_{\bf k}s_{\bf k}^{\phantom{\dagger}}
        n_{\bf k}^{\phantom{\dagger}}+
        c^*_{-\bf k} s_{-\bf k}^{\phantom{\dagger}}
        (n_{-\bf k}^{\phantom{\dagger}}  + 1) |^2 }
\over \displaystyle\phantom{|} {n_1({\bf k}) \, n_1({- \bf k})} }
        \left | \int dt F(t) \exp \left [-i 
	 (\omega_{\bf k} + \omega_{-{\bf k}})t \right ] \right |^2 .
         \label{e:c2bs}
\end{equation}
For a typical exponential decay,
$F(t)  = \theta(t-t_0)\Gamma \exp[-\Gamma (t-t_0)] $
with $ \delta t = \hbar/\Gamma = 2$ fm/c,  
we show the BBC for the $\phi$ mesons in Fig. 1,
which shows that the BBC survives the
suppression with a strength as large as 2-3.

\section{RESULTS FOR INHOMOGENEOUS SYSTEMS}
Following Ref.\cite{sm}, we divide the inhomogeneous fluid into
cells labeled $i$.
In each cell, it is assumed that the field can be expanded with 
creation and annihilation operators, 
and that $H_i$ is diagonalized by a local Bogoliubov transformation,
that implies
\begin{eqnarray}
G_c(1,2) & = & 
        {\displaystyle\phantom{|} 1 \over (2 \pi)^3 }
        \int d^4\sigma_{\mu}(x) K_{12}^{\mu} e^{i q_{12} x}
         \left[|c_{1,2}|^2   n_{1,2} + 
                | s_{-1,-2}|^2  (n_{-1,-2} + 1) \right],\label{e:gc} \\
G_s(1,2) & = & 
        {\displaystyle\phantom{|} 1 \over (2 \pi)^3 }
        \int d^4\sigma_{\mu}(x) K_{12}^{\mu} e^{i 2 K_{12} x}
        \left[ s^*_{-1,2} c_{2,-1} n_{-1,2}  +
                c_{1,-2} s^*_{-2,1} (n_{1,-2} + 1)
                \right]. 
        \label{e:gd}\label{e:gs}
\end{eqnarray}
Here $d^4\sigma^{\mu}(x) = d^3\Sigma^{\mu}(x;\tau_f)\, F(\tau_f)  
d\tau_f $ is the product of the normal-oriented
volume element depending parametrically on $\tau_f$ (the freeze-out 
hypersurface parameter) and the
invariant distribution of that parameter $F(\tau_f)$.
The other variables are defined as follows:
\begin{eqnarray}
n_{i,j} (x) & = & 1 / \left[ 
        \exp[ (K^{* \mu}_{i,j}(x) u_{\mu}(x) -  \mu(x)) / T(x) ] 
- 1 \right],\\
r(i,j,x) & = & \frac{1}{2}\log
        \left[( K^{\mu}_{i,j}(x) u_\mu (x))/ 
        (K^{* \mu}_{i,j}(x) u_\mu(x) ) \right],
        \label{e:rxk}\\
c_{i,j} & =  & \cosh[r({i,j},x)] , \quad 
s_{i,j} \, = \, \sinh[r({i,j},x)],
\end{eqnarray} 
where $i,j = \pm 1,\pm 2$  and the mean and the relative momenta
for the $a$($b$)-quanta are defined as  
$K^{(*) \mu}_{i,j}(x) = [k^{(*) \mu}_i(x) + k^{(*) \mu}_j(x) ]/2 $ and
$q^{\mu}_{i,j} = k^{\mu}_i - k^{\mu}_j , $ respectively.
The local hydrodynamical flow field is denoted by $u^{\mu}(x)$.
See ref. ~\cite{acg} for further details.
The correlation function in the presence of local
squeezing is given by
\begin{equation}
C_2({\bf k}_1,{\bf k}_2) = 
        1 
        + 
        {|G_c(1,2) |^2 \over G_c(1,1) G_c(2,2) }
        + {|G_s(1,2)|^2\over G_c(1,1) G_c(2,2) } . 
\end{equation}
As the Bogoliubov transformation always mixes particles
with anti-particles, the above expression holds only for the case
where particles are equivalent to their anti-particles,
e.g. the $\phi$ meson and $\pi^0$.
However, the extension to the case where particles and their
anti-particles are different is straightforward;
correlations between particles and anti-particles such
as $\pi^+$ and $\pi^-$, $K^+$ and $K^-$, and so forth, appear~\cite{acg}.
Fig. 2 illustrates the novel character of 
BBC for two identical bosons caused
by medium mass-modifications, 
along with the familiar Bose-Einstein or HBT correlations on the
diagonal of the $({\bf k}_1 , {\bf k}_2)$ plane.

\section{SUMMARY}
The theory of particle correlations and spectra
for bosons with in-medium mass-shifts  predicts
the existence of back-to-back correlations of 
$\phi\phi$, $K^+K^-$, $\pi^0 \pi^0$ and $\pi^+ \pi^-$ pairs
that  could be searched for at CERN SPS and
upcoming RHIC BNL heavy ion experiments \cite{lz}.
Surprisingly, such  novel back-to-back correlations could be as large as
the well-known HBT correlations, surviving large
finite time suppression factors.

\vspace*{-0.2cm}
\begin{figure}[hbt]
\begin{minipage}[t]{76mm}
\epsfig{file=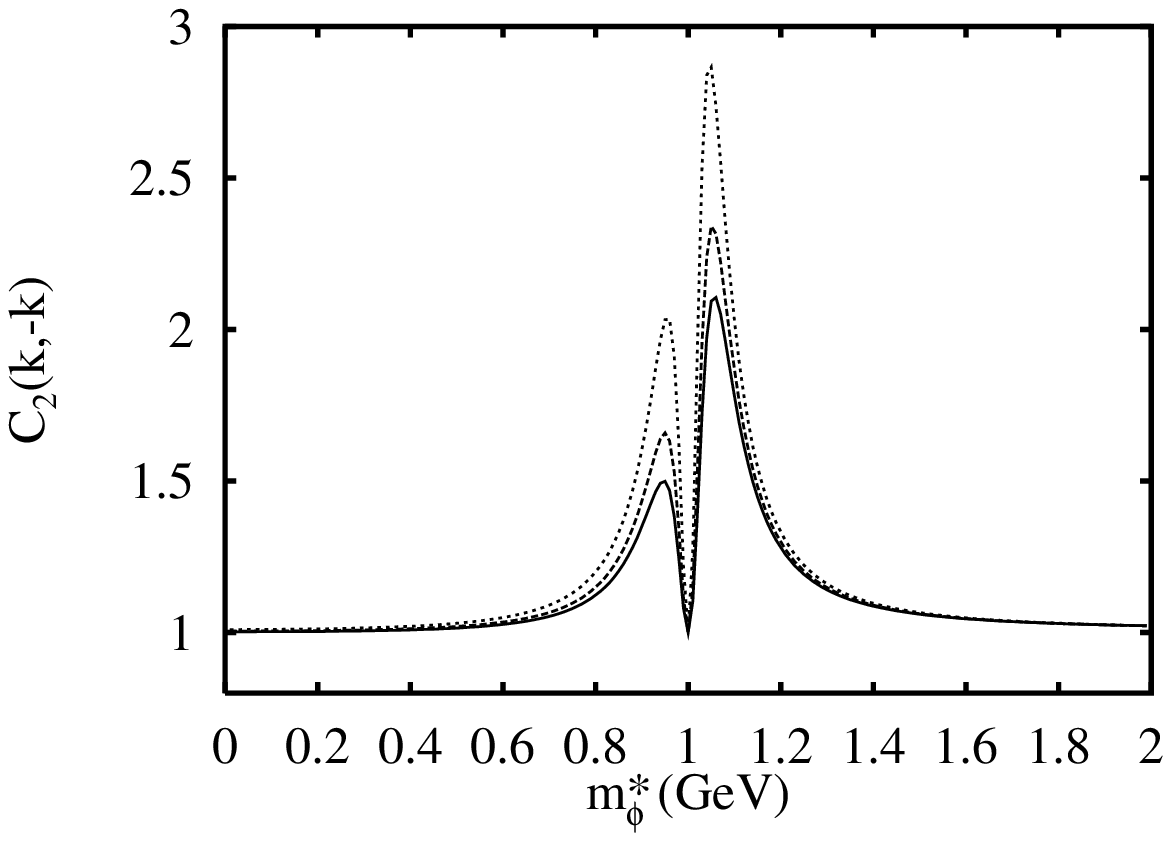,height=2.05in,angle=0}
\vspace*{-0.7cm}
\caption{Dependence of the back-to-back correlations (BBC) 
        on the medium modified $\phi$ meson mass, $m^*_{\phi}$,
        for $T = 140$ MeV, $\mu = 0$, and $\delta t = 2$ fm/c,
        where solid, dashed and dotted line 
        stands for $|{\bf k}| = 0 $, 300 and 500 MeV, respectively.
        The magnitude of the BBC is large
        in spite of the finite time  suppression.
}
\end{minipage}
\hspace*{0.6cm}
\begin{minipage}[t]{76mm}
\epsfig{file=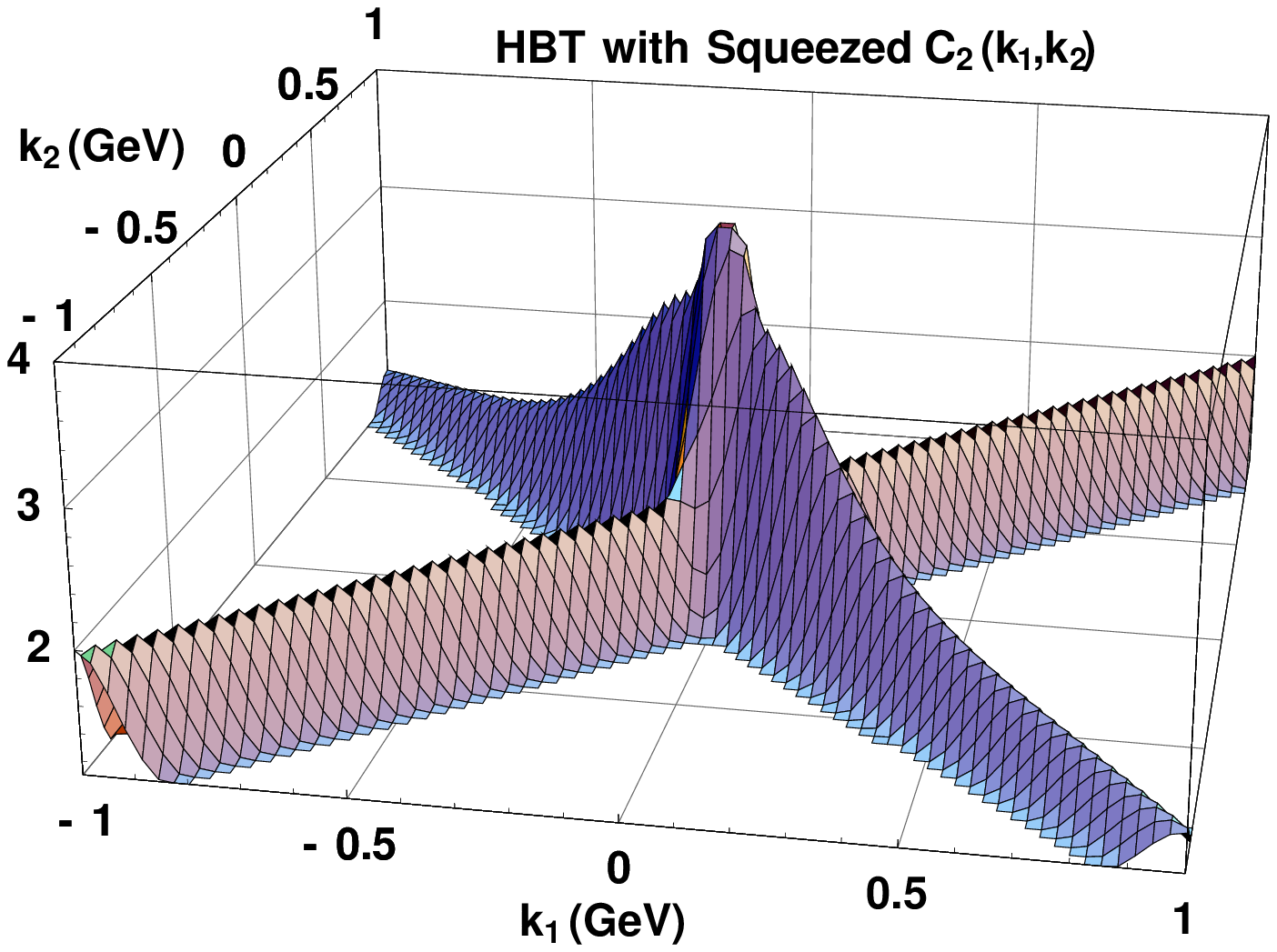,height=2.05in,angle=0}
\vspace*{-0.7cm}
\caption{Schematic illustration of the new kind of correlations
        for mass shifted $\pi^0$ pairs, assuming $T = 140 $ MeV, 
	$\delta t \!=\! 0$
        $G_c \!\sim\! \exp[-q_{12}^2 R_G^2/2 ], 
G_s \!\sim\! \exp[-2 K_{12}^2 R_G^2]$, 
        with $R_G \! =\! 2$ fm. 
The fall of the BBC for increasing
values of $ |{\bf k} |$ is controlled here by a momentum-dependent mass-shift,
$m^*_{\pi} \! =\! m_{\pi} [1 + \exp( - {\bf k}^2 / \Lambda_s^2) ] $
with $\Lambda_s = 325 $ MeV.
}
\end{minipage}
\end{figure}

\end{document}